\pdfoutput=1
\documentclass[12pt,a4paper]{article}

\usepackage{ifthen} 
\newboolean{pdflatex}
\setboolean{pdflatex}{true} 

\newboolean{articletitles}
\setboolean{articletitles}{true} 

\newboolean{uprightparticles}
\setboolean{uprightparticles}{false} 

\def\paperauthors{LHCb collaboration} 
\def\paperasciititle{First measurement of time-dependent  CP violation in the flavor-changing neutral-current decay  B0 to KS0 mumu } 
\def\papertitle{First measurement of time-dependent  \CP violation in the flavor-changing neutral-current decay  \mbox{$\Bd\to K_{\rm S}^{0}\mu^{+}\mu^{-}$ } } 
\def\paperkeywords{{High Energy Physics}, {LHCb}} 
\def\papercopyright{\the\year\ CERN for the benefit of the LHCb collaboration} 
\def\paperlicence{CC BY 4.0 licence}
\def\paperlicenceurl{https://creativecommons.org/licenses/by/4.0/}

\newif\ifEnableSectionTOCLinks
\EnableSectionTOCLinksfalse 

\usepackage[top=1in, bottom=1.25in, left=1in, right=1in]{geometry}

\usepackage{microtype}
\usepackage{lineno}  
\usepackage{xspace} 
\usepackage{caption} 

\usepackage{graphicx}  
\usepackage{color}
\usepackage{colortbl}
\graphicspath{{./figs/}} 

\usepackage{amsmath} 
\usepackage{amssymb}
\usepackage{amsfonts}
\usepackage{upgreek} 

\newcommand*\patchAmsMathEnvironmentForLineno[1]{%
\expandafter\let\csname old#1\expandafter\endcsname\csname #1\endcsname
\expandafter\let\csname oldend#1\expandafter\endcsname\csname
end#1\endcsname
 \renewenvironment{#1}%
   {\linenomath\csname old#1\endcsname}%
   {\csname oldend#1\endcsname\endlinenomath}%
}
\newcommand*\patchBothAmsMathEnvironmentsForLineno[1]{%
  \patchAmsMathEnvironmentForLineno{#1}%
  \patchAmsMathEnvironmentForLineno{#1*}%
}
\AtBeginDocument{%
\patchBothAmsMathEnvironmentsForLineno{equation}%
\patchBothAmsMathEnvironmentsForLineno{align}%
\patchBothAmsMathEnvironmentsForLineno{flalign}%
\patchBothAmsMathEnvironmentsForLineno{alignat}%
\patchBothAmsMathEnvironmentsForLineno{gather}%
\patchBothAmsMathEnvironmentsForLineno{multline}%
\patchBothAmsMathEnvironmentsForLineno{eqnarray}%
}

\usepackage[pdftex,
            pdfauthor={\paperauthors},
            pdftitle={\paperasciititle},
            pdfkeywords={\paperkeywords}]{hyperref}
\usepackage{hyperxmp}
\hypersetup{
    pdfcopyright={Copyright (C) \papercopyright},
    pdflicenseurl={\paperlicenceurl}
}

\usepackage[colorinlistoftodos,textsize=scriptsize]{todonotes}

\usepackage[bottom,flushmargin,hang,multiple]{footmisc}

\usepackage[all]{hypcap} 

\usepackage{xspace} 
\usepackage{upgreek}

\def\lhcb   {\mbox{LHCb}\xspace}

\def\MagUp {\mbox{\em Mag\kern -0.05em Up}\xspace}

\def\sPlot{\mbox{\em sPlot}\xspace}

\ifthenelse{\boolean{uprightparticles}}%
{

 \def\Pmu         {\ensuremath{\upmu}\xspace}

 \def\Ppi         {\ensuremath{\uppi}\xspace}

 \def\Ppsi        {\ensuremath{\uppsi}\xspace}

 \def\PDelta      {\ensuremath{\Delta}\xspace}                 
 \def\PXi         {\ensuremath{\Xi}\xspace}                 
 \def\PLambda     {\ensuremath{\Lambda}\xspace}                 
 \def\PSigma      {\ensuremath{\Sigma}\xspace}                 
 \def\POmega      {\ensuremath{\Omega}\xspace}                 
 \def\PUpsilon    {\ensuremath{\Upsilon}\xspace}
 \let\oldPi\Pi
 \def\PPi         {\ensuremath{\oldPi}\xspace}

 \def\PB      {\ensuremath{\mathrm{B}}\xspace}                 
 \def\PD      {\ensuremath{\mathrm{D}}\xspace}                 
 \def\PJ      {\ensuremath{\mathrm{J}}\xspace}                 
 \def\PK      {\ensuremath{\mathrm{K}}\xspace}                 
 \def\Pb      {\ensuremath{\mathrm{b}}\xspace}                 
                  
 \def\Pd      {\ensuremath{\mathrm{d}}\xspace}                 

 \def\Ps      {\ensuremath{\mathrm{s}}\xspace}

 \def\thebaroffset{0.0em}
}
{

 \def\Pmu         {\ensuremath{\mu}\xspace}

 \def\Ppi         {\ensuremath{\pi}\xspace}

 \def\Ppsi        {\ensuremath{\psi}\xspace}                 
                  
 \mathchardef\PDelta="7101
 \mathchardef\PXi="7104
 \mathchardef\PLambda="7103
 \mathchardef\PSigma="7106
 \mathchardef\POmega="710A
 \mathchardef\PUpsilon="7107
 \mathchardef\PPi="7105
 \def\PB      {\ensuremath{B}\xspace}                 
 \def\PD      {\ensuremath{D}\xspace}                 
 \def\PJ      {\ensuremath{J}\xspace}                 
 \def\PK      {\ensuremath{K}\xspace}                 
 \def\Pb      {\ensuremath{b}\xspace}                 
                  
 \def\Pd      {\ensuremath{d}\xspace}                 

 \def\Ps      {\ensuremath{s}\xspace}

 \def\thebaroffset{0.18em}
}
\newcommand{\offsetoverline}[2][\thebaroffset]{\kern #1\overline{\kern -#1 #2}}%

\makeatletter
\ifcase \@ptsize \relax
  \newcommand{\miniscule}{\@setfontsize\miniscule{4}{5}}
\or
  \newcommand{\miniscule}{\@setfontsize\miniscule{5}{6}}
\or
  \newcommand{\miniscule}{\@setfontsize\miniscule{5}{6}}
\fi
\makeatother

\DeclareRobustCommand{\optbar}[1]{\shortstack{{\miniscule (\rule[.5ex]{1.25em}{.18mm})}
  \\ [-.7ex] $#1$}}

\def\mup        {{\ensuremath{\Pmu^+}}\xspace}
\def\mun        {{\ensuremath{\Pmu^-}}\xspace} 

\def\dquark    {{\ensuremath{\Pd}}\xspace}

\def\squark    {{\ensuremath{\Ps}}\xspace}

\def\bquark    {{\ensuremath{\Pb}}\xspace}

\def\pion   {{\ensuremath{\Ppi}}\xspace}

\def\pip    {{\ensuremath{\pion^+}}\xspace}
\def\pim    {{\ensuremath{\pion^-}}\xspace}

\def\kaon    {{\ensuremath{\PK}}\xspace}

\def\KorKbar {\kern \thebaroffset\optbar{\kern -\thebaroffset \PK}{}\xspace}

\def\Kp      {{\ensuremath{\kaon^+}}\xspace}

\def\KS      {{\ensuremath{\kaon^0_{\mathrm{S}}}}\xspace}

\def\Kstarz  {{\ensuremath{\kaon^{*0}}}\xspace}

\def\Kstar   {{\ensuremath{\kaon^*}}\xspace}

\def\D       {{\ensuremath{\PD}}\xspace}

\def\DorDbar {\kern \thebaroffset\optbar{\kern -\thebaroffset \PD}\xspace}

\def\Dp      {{\ensuremath{\D^+}}\xspace}
\def\Dm      {{\ensuremath{\D^-}}\xspace}

\def\DpDm    {\ensuremath{\Dp {\kern -0.16em \Dm}}\xspace}

\def\B       {{\ensuremath{\PB}}\xspace}
\def\Bbar    {{\ensuremath{\offsetoverline{\PB}}}\xspace}

\def\BorBbar {\kern \thebaroffset\optbar{\kern -\thebaroffset \PB}\xspace}
\def\Bz      {{\ensuremath{\B^0}}\xspace}
\def\Bzb     {{\ensuremath{\Bbar{}^0}}\xspace}
\def\Bd      {{\ensuremath{\B^0}}\xspace}
\def\Bdb     {{\ensuremath{\Bbar{}^0}}\xspace}
\def\BdorBdbar {\kern \thebaroffset\optbar{\kern -\thebaroffset \Bd}\xspace}
\def\Bu      {{\ensuremath{\B^+}}\xspace}

\def\Bs      {{\ensuremath{\B^0_\squark}}\xspace}

\def\BsorBsbar {\kern \thebaroffset\optbar{\kern -\thebaroffset \Bs}\xspace}

\def\jpsi     {{\ensuremath{{\PJ\mskip -3mu/\mskip -2mu\Ppsi}}}\xspace}
\def\psitwos  {{\ensuremath{\Ppsi{(2S)}}}\xspace}

\def\Y#1S{\ensuremath{\PUpsilon{(#1S)}}\xspace}

\def\Lz          {{\ensuremath{\PLambda}}\xspace}

\def\LorLbar     {\kern \thebaroffset\optbar{\kern -\thebaroffset \PLambda}\xspace}

\def\Lb           {{\ensuremath{\Lz^0_\bquark}}\xspace}

\newcommand{\decay}[2]{\ensuremath{\mathinner{#1\!\to #2}}\xspace}

\def\to                 {\ensuremath{\rightarrow}\xspace}

\def\CP                {{\ensuremath{C\!P}}\xspace}

\newcommand{\dmd}{{\ensuremath{\Delta m_{\dquark}}}\xspace}

\newcommand{\etag}{{\ensuremath{\varepsilon_{\mathrm{tag}}}}\xspace}

\def\AT#1     {\ensuremath{A_{\mathrm{T}}^{#1}}\xspace}           

\def\C#1      {\ensuremath{\mathcal{C}_{#1}}\xspace}                       
\def\Cp#1     {\ensuremath{\mathcal{C}_{#1}^{'}}\xspace}                    
\def\Ceff#1   {\ensuremath{\mathcal{C}_{#1}^{\mathrm{(eff)}}}\xspace}        
\def\Cpeff#1  {\ensuremath{\mathcal{C}_{#1}^{'\mathrm{(eff)}}}\xspace}       
\def\Ope#1    {\ensuremath{\mathcal{O}_{#1}}\xspace}                       
\def\Opep#1   {\ensuremath{\mathcal{O}_{#1}^{'}}\xspace}                    

\newcommand{\aunit}[1]{\ensuremath{\text{\,#1}}}       

\newcommand{\tev}{\aunit{Te\kern -0.1em V}\xspace}
\newcommand{\gev}{\aunit{Ge\kern -0.1em V}\xspace}
\newcommand{\mev}{\aunit{Me\kern -0.1em V}\xspace}
\newcommand{\kev}{\aunit{ke\kern -0.1em V}\xspace}
\newcommand{\ev}{\aunit{e\kern -0.1em V}\xspace}
 
\newcommand{\mevc}{\ensuremath{\aunit{Me\kern -0.1em V\!/}c}\xspace}
\newcommand{\gevc}{\ensuremath{\aunit{Ge\kern -0.1em V\!/}c}\xspace}
\newcommand{\mevcc}{\ensuremath{\aunit{Me\kern -0.1em V\!/}c^2}\xspace}
\newcommand{\gevcc}{\ensuremath{\aunit{Ge\kern -0.1em V\!/}c^2}\xspace}
\newcommand{\gevgevcccc}{\ensuremath{\gev^2\!/c^4}\xspace} 

\def\fb   {\ensuremath{\aunit{fb}}\xspace}
\def\invfb   {\ensuremath{\fb^{-1}}\xspace}

\def\ps   {\ensuremath{\aunit{ps}}\xspace}
\def\fs   {\aunit{fs}}

\def\invps{\ensuremath{\ps^{-1}}\xspace}

\def\gsim{{~\raise.15em\hbox{$>$}\kern-.85em
          \lower.35em\hbox{$\sim$}~}\xspace}
\def\lsim{{~\raise.15em\hbox{$<$}\kern-.85em
          \lower.35em\hbox{$\sim$}~}\xspace}

\def\sPlot{\mbox{\em sPlot}\xspace}

\def\pt         {\ensuremath{p_{\mathrm{T}}}\xspace}

\def\tell1  {TELL1\xspace}
\def\ukl1   {UKL1\xspace}

\newcommand{\lhcborcid}[1]{\href{https://orcid.org/#1}{\hspace*{0.1em}\raisebox{-0.45ex}{\includegraphics[width=1em]{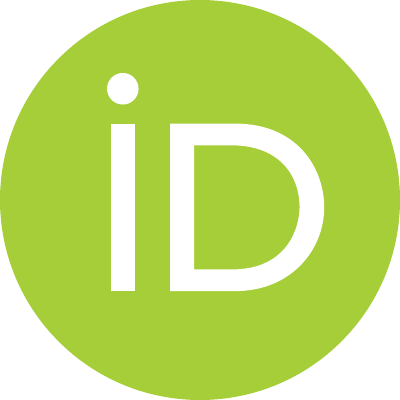}}}}

\hypersetup{
  colorlinks   = true, 
  urlcolor     = blue, 
  linkcolor    = blue, 
  citecolor    = red   
}

\ifEnableSectionTOCLinks
    \usepackage[explicit]{titlesec} 
    
    \let\oldcontentsline\contentsline
    \renewcommand

    \titleformat{\section}{\normalfont\Large\bf}{\hyperlink{tocsection.\thesection}{{\thesection} \parbox[t]{\dimexpr\textwidth-1pc}{#1}}}{1pc}{}

    \titleformat{\subsection}{\normalfont\bf}{\hyperlink{tocsubsection.\thesubsection}{{\thesubsection} \parbox[t]{\dimexpr\textwidth-1pc}{#1}}}{1pc}{}

    \titleformat{name=\section,numberless}[display]{}{}{0pt}{\normalfont\Huge\bfseries #1}
\fi

\usepackage{cite} 
\usepackage{LHCb/mciteplus}

\usepackage{longtable} 

\begin{document}

\renewcommand{\thefootnote}{\fnsymbol{footnote}}
\setcounter{footnote}{1}


\begin{titlepage}
\pagenumbering{roman}

\vspace*{-1.5cm}
\centerline{\large EUROPEAN ORGANIZATION FOR NUCLEAR RESEARCH (CERN)}
\vspace*{1.5cm}
\noindent
\begin{tabular*}{\linewidth}{lc@{\extracolsep{\fill}}r@{\extracolsep{0pt}}}
\ifthenelse{\boolean{pdflatex}}
{\vspace*{-1.5cm}\mbox{\!\!\!\includegraphics[width=.14\textwidth]{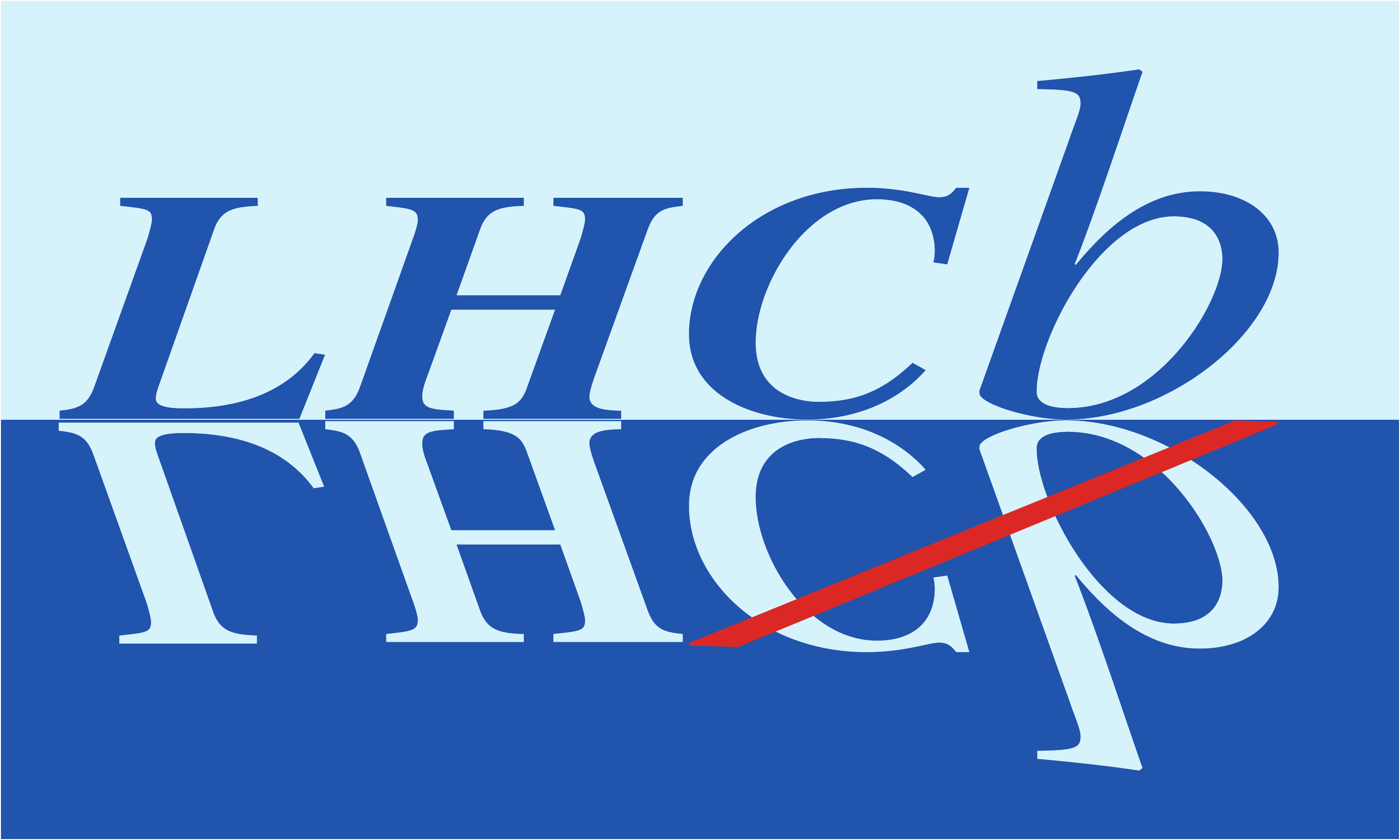}} & &}%
{\vspace*{-1.2cm}\mbox{\!\!\!\includegraphics[width=.12\textwidth]{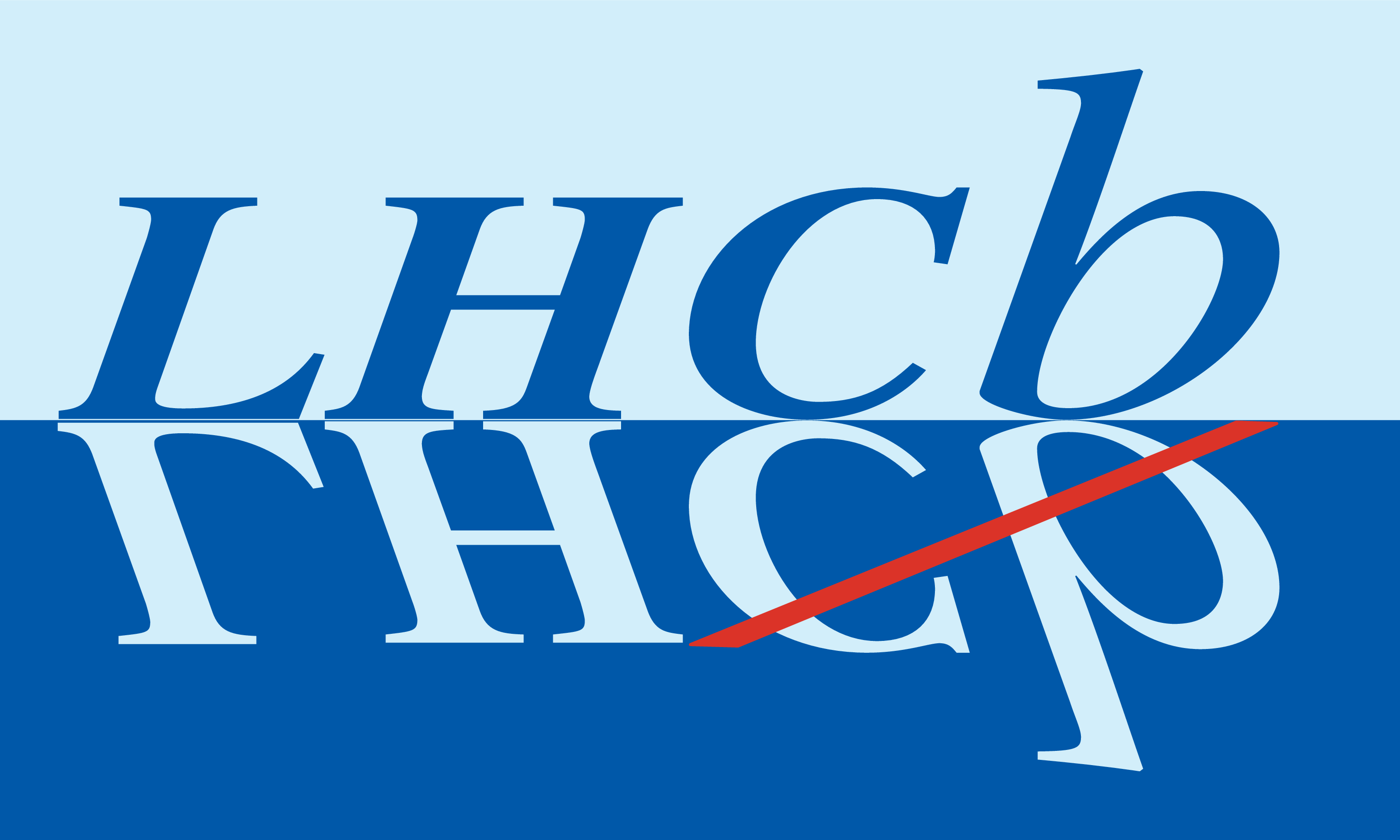}} & &}%
\\
 & & CERN-EP-2026-037 \\  
 & & LHCb-PAPER-2025-062 \\  
 & &  15 March 2026 \\ 
 & & \\
\end{tabular*}

\vspace*{2.0cm}

{\normalfont\bfseries\boldmath\huge
\begin{center}
  \papertitle 
\end{center}
}

\vspace*{2.0cm}

\begin{center}
\paperauthors\footnote{Authors are listed at the end of this paper.}
\end{center}

\vspace{\fill}

\begin{abstract}
  \noindent
 A flavor-tagged time-dependent  analysis of  \mbox{$\Bd\to K_{\rm S}^{0}\mu^{+}\mu^{-}$ } decays is performed across the full dimuon mass range excluding the \jpsi and $\psi(2S)$ resonance regions. The analysis uses  proton-proton collision data
 collected by the \lhcb experiment in 2011--2018
 at center-of-mass energies of 7, 8 and 13\tev, corresponding to an integrated luminosity of 9\invfb.
 The \CP violation  parameters  are determined to be  
\begin{eqnarray}
 &C=-0.13 \pm 0.32  \pm 0.04 \,,\nonumber \\
 &S=+0.82\pm 0.29  \pm 0.05 \,,\nonumber
\end{eqnarray}
  where
 the first uncertainties
are statistical and the second are systematic.
 The results are consistent with   
 the Standard Model prediction.
 This is the first experimental study of time-dependent  \CP violation in \mbox{$b\rightarrow s \ell^{+} \ell^{-}$}  processes.
  
\end{abstract}

\vspace*{2.0cm}

\begin{center}
  Published on
  Phys.~Rev.~Lett. 
\end{center}

\vspace{\fill}

{\footnotesize 
\centerline{\copyright~\papercopyright. \href{\paperlicenceurl}{\paperlicence}.}}
\vspace*{2mm}

\end{titlepage}


\newpage
\setcounter{page}{2}
\mbox{~}
%
%
%
%


\renewcommand{\thefootnote}{\arabic{footnote}}
\setcounter{footnote}{0}


\cleardoublepage


\pagestyle{plain} 
\setcounter{page}{1}
\pagenumbering{arabic}


Flavor-changing neutral-current decays~\cite{FCNC2017}, such as the rare \decay{b}{s \ell^+ \ell^-} transition, serve as powerful and sensitive probes for physics beyond the Standard Model (SM). Their inherent suppression within the SM makes any observed deviation in their decay properties 
a clear signature of new physics.  
Since beginning of operations,  the \lhcb experiment has performed extensive studies of  a series of \decay{b}{s \ell^+\ell^-} processes, which have revealed a compelling, albeit evolving, picture of anomalies.Most notably,  the measurements of branching fractions and angular observables in several semileptonic $B$-meson decays performed by the \lhcb, ATLAS, CMS, Belle, and BaBar collaborations~\cite{LHCb-PAPER-2021-022,LHCb-PAPER-2021-014,LHCb-PAPER-2018-029,LHCb-PAPER-2022-050,LHCb-PAPER-2014-049,LHCb-PAPER-2020-041,LHCb-PAPER-2017-001,LHCb-PAPER-2020-002,LHCb-PAPER-2025-041,ATLAS:2018gqc,CMS:2017rzx,Belle:2016fev,BaBar:2015wkg}  show a coherent and distinct pattern of tensions with the SM, which could be explained by new-physics contributions to the effective four-fermion \decay{b}{s \ell^+\ell^-} couplings~\cite{Alguero2023,Allanach:2023uxz,Capdevila:2023yhq}.

These persistent anomalies motivate the use of complementary experimental probes to better understand the underlying dynamics.
 In particular, time-dependent 
 analyses of neutral $B$-meson decays provide theoretically clean observables to explore new-physics phenomena in \decay{b}{s} transitions~\cite{Jager2013,Descotes_Genon_2021,Descotes-Genon2015,addref1,addref2,addref3,addref4,addref5,addref6,addref7}.  
 The \CP asymmetry arising from the interference between  the 
$\Bz$--$\Bzb$ mixing and decay amplitudes is sensitive to the imaginary parts of the Wilson coefficients governing these processes,  
yielding constraints on new \CP-violating phases that are complementary to, 
 and in many cases more direct and robust than, those obtained from the time-integrated or time-independent 
 decay rates, angular distributions and  \CP asymmetries  in  decays such as $\Bd \to K^{*}(892)^0 \mup\mun$~\cite{Descotes-Genon2011,Fleischer:2025ucq,LHCb-PAPER-2024-011}.  The power of the time-dependent  approach is well established in hadronic \decay{b}{s q \bar{q}} $(q=d, s)$ channels, such as \decay{\Bz}{\eta' \KS} and \decay{\Bz}{\phi(1020) \KS}~\cite{PhysRevLett804625,GROSSMAN1997241}.\footnote{The inclusion of charge-conjugate processes is implied throughout this Letter unless otherwise stated.}
 The decay \decay{\Bd}{\KS \mup\mun} thus   enables  the first time-dependent \CP violation measurement in a \decay{b}{s \ell^+ \ell^-} process, opening a crucial new window to \CP-violating new physics in these rare modes.

In general, the $\KS\mup\mun$ final state is an admixture of \CP-even and odd eigenstates,  thus a full time-dependent angular analysis  is required to perform  the most complete test of \CP violation~\cite{Descotes_Genon_2021,Alguero:2021yus}.  Given the  limited statistics of the data sample considered in this Letter, this study 
focuses on measuring the overall time-dependent \CP asymmetry after integrating over the lepton helicity angle in the dimuon rest frame. 
The time-dependent  \CP asymmetry between the decay rates of  initial \Bzb and \Bz mesons to the $\KS\mup\mun$  final state is defined as 
\begin{equation}
\label{eq1}
    \mathcal{A}^{\CP}(t) \equiv \frac{\Gamma(\decay{\Bzb(t)}{\KS\mup\mun}) - \Gamma(\decay{\Bz(t)}{\KS\mup\mun})}{\decay{\Gamma(\Bzb(t)}{\KS\mup\mun}) + \Gamma(\decay{\Bz(t)}{\KS\mup\mun})}\simeq S \sin(\Delta m_d t) - C \cos(\Delta m_d t),
\end{equation}
where $t$ is the time interval between the production and decay of the \Bzb or \Bz meson in its rest frame, $\Delta m_d$ is the $\Bz$--$\Bzb$ oscillation frequency, and $\Gamma$ is the decay rate for corresponding processes. The decay-width difference of the two $\Bz$ mass eigenstates, $\Delta\Gamma_d$ is assumed to be zero in the equation.
The parameters $S$ and $C$ quantify the mixing-induced and direct \CP violation, respectively. 
 In the SM, the \decay{\Bd}{\KS\mup\mun} decay proceeds dominantly to a \CP-odd state  
and  the expectation is 
$C\approx 0$ and 
$S =\sqrt{(1-C^2)} \sin(2\beta + \Delta\phi_d ) \approx \sin (2\beta)$. Here  $\beta$ is the CKM angle measured in \decay{\Bd}{\jpsi \KS} decays~\cite{LHCb-PAPER-2023-013}, and $\Delta\phi_d$ is a small phase shift due to  the interference of the  \Bz mixing and \decay{b}{s} decay amplitudes.
For  the dimuon mass-squared range $1.1<q^2< 6.0$\gevgevcccc,
the mixing-induced \CP asymmetry  is explicitly predicted to be 
 $S=0.72\pm 0.02$ in the SM~\cite{Fleischer2023}.
However, various new-physics models with new interactions or 
additional weak phases can  lead to  a significant deviation of  $S$ and $C$ from their SM predictions~\cite{PRD107075}.

This Letter reports the first measurement of the  \CP-violating parameters  in  \decay{\Bd}{\KS \mup\mun} decays using $pp$ collision data  collected with the \lhcb  detector in 2011--2012 (Run~1) and 2015--2018 (Run~2), 
corresponding to an integrated luminosity of 9\invfb. 
The \CP-violating observables $C$ and $S$ are measured from a time-dependent  analysis in the  full range of dimuon masses, $q^2$, excluding  
 the $\jpsi$ ($8.0<q^2<11.0$\gevgevcccc)  and $\psi(2S)$  ($15.0<q^2<17.5$\gevgevcccc) regions.  
The control channel  \decay{\Bd}{\jpsi(\to \mup\mun ) \KS} is used to  study the  difference between the simulation  and  data and to validate the fit procedure.

The \lhcb detector~\cite{LHCb-DP-2008-001,LHCb-DP-2014-002} is a single-arm forward spectrometer covering the pseudorapidity range $2 < \eta < 5$, designed for the study of particles containing $b$ or $c$ quarks. It employs a dipole magnet with periodically reversed polarity for momentum determination. The detector used to collect the
data analyzed in this paper includes a high-precision tracking system consisting of silicon-strip~\cite{LHCb-DP-2014-001,LHCb-DP-2017-001} and straw drift-tube~\cite{LHCb-DP-2017-001} detectors. Different types of charged hadrons are distinguished using information from two ring-imaging Cherenkov detectors~\cite{LHCb-DP-2013-003}.  Muons are identified by a
 system composed of alternating layers of iron and multiwire proportional chambers, achieving high efficiency and excellent momentum resolution~\cite{LHCb-DP-2012-002,LHCb-DP-2013-001}.

 The simulated samples used to study the decay properties of the signal, control, and background channels are
generated using the software described in Refs.~\cite{Sjostrand:2007gs,*Sjostrand:2006za,Lange:2001uf,Golonka:2005pn,Allison:2006ve, *Agostinelli:2002hh,LHCb-PROC-2011-006,LHCb-PROC-2010-056,LHCb-PUB-2016-021}.
        Simulated signal  decays are generated according to the theoretical calculation for \decay{b}{s} transitions in Ref.~\cite{Ball:2004rg}. The data-simulation difference is studied using the control channel \decay{\Bd}{\jpsi \KS}, and applied to the signal simulation to match the \Bz kinematic distributions and the particle identification (PID) performance in data.~\cite{LHCb-DP-2018-001,LHCb-DP-2013-001,LHCb-PUB-2016-021}.

The online event selection
is performed by a trigger~\cite{LHCb-DP-2012-004},
which consists of a hardware stage followed by a two-stage software trigger.  The hardware trigger  for this study
     employs information from the  muon system.
Subsequently, the  software trigger imposes requirements on the muon
transverse momentum and track reconstruction quality, or on the output of  a multibody  topological $b$-hadron trigger~\cite{LHCb-PROC-2015-018}.
Triggered data further undergo a centralized, offline processing step
 across the entire \lhcb physics program~\cite{Stripping}.

In the offline analysis, $\KS$ candidates are reconstructed from two oppositely charged tracks  considered as pions. They are categorized according to the position of the $\KS$ decay vertex.
For $\KS$ mesons that decay inside the vertex locator (VELO), the resulting tracks are classified as long tracks and are reconstructed using hits from both the VELO and the main tracking stations, thereby providing the best momentum resolution. For particles that decay outside the VELO, the tracks are classified as downstream tracks; these are reconstructed using only hits from the tracking stations. The  mass of the $\pip\pim$ pair is required to be within $20\mevcc$ of the known $\KS$ mass~\cite{PDG2024}. 
These \KS candidates are then combined with a pair of oppositely charged muons to form  \Bd candidates.  
All tracks and vertices are required to have a good fit quality. 
Furthermore, PID criteria are applied to ensure that the charged tracks are consistent with the hypothesis of being pions or muons, as required by the decay chain. 
The PID observables of the charged tracks are calculated  using multivariate techniques to combine information from different subsystems~\cite{LHCb-DP-2013-001,LHCb-PUB-2016-021,LHCb-DP-2018-001}.
A kinematic fit is performed to improve the \Bd  mass resolution by constraining the mass of the $\pip\pim$  pair to the known  \KS mass\cite{PDG2024}.

This analysis considers specific backgrounds that can contaminate the \decay{\Bd}{\KS \mup\mun} signal sample. These include \decay{\Lb}{\Lz (\to p\pim)\mup\mun} decays where the proton is misidentified as a pion, and \decay{\Bd}{\Dm (\to \KS\pim)\pip} and \decay{\Bd}{\KS \pip\pim}  decays where two pions are misidentified as muons. These backgrounds are highly suppressed  through mass vetos and PID requirements. Evaluation using simulation samples confirms that their residual contamination to the final signal sample is negligible after these requirements.
The $q^2$ value is required to be outside the two defined charmonium regions, except when selecting the control channel  \decay{\Bd}{\jpsi(\to \mup\mun ) \KS}.

To further suppress the combinatorial background arising from the random combination of unrelated tracks, a multivariate analysis based on a gradient boosted decision tree (BDTG) technique is carried out~\cite{Friedman2001GreedyFA}. The BDTG classifier is trained using simulated \decay{\Bd}{ \KS \mup \mun} signal decays  and  data candidates in the high \KS\mup\mun  mass sideband of  
[5700, 6000]\mevcc  to represent the background. 
The BDTG classifier incorporates a set of kinematic and topological variables, such as the transverse momentum (\pt) and $\eta$ of the \Bd and \KS candidates, their impact parameters  with respect to the primary vertex (PV), and PID variables of the final state particles. These inputs show good agreement between the background-subtracted data and simulation in both the signal and control channel \decay{\Bd}{\jpsi \KS}.  
The  requirement on the BDTG output is chosen to maximize the expected signal significance.

Eight categories of  \decay{\Bz}{\KS\mup\mun} candidates are selected using separate BDTG classifiers, corresponding to the combinations of  four data-taking periods with roughly equal signal yields (2011--2012, 2015--2016, 2017, 2018) and two \KS categories (with 
long or downstream track types). An unbinned extended maximum-likelihood fit is performed to the \Bd mass distribution  to determine the signal and background yields in  each category. The \Bd signal is modeled by a Crystal Ball function~\cite{Skwarnicki:1986xj}, with tail parameters fixed from simulation and the width and peak position allowed to vary in the fit.  
The combinatorial background
is modeled with an exponential function. 

\begin{figure}[tb]
\centering
   \includegraphics[width=0.5\linewidth]{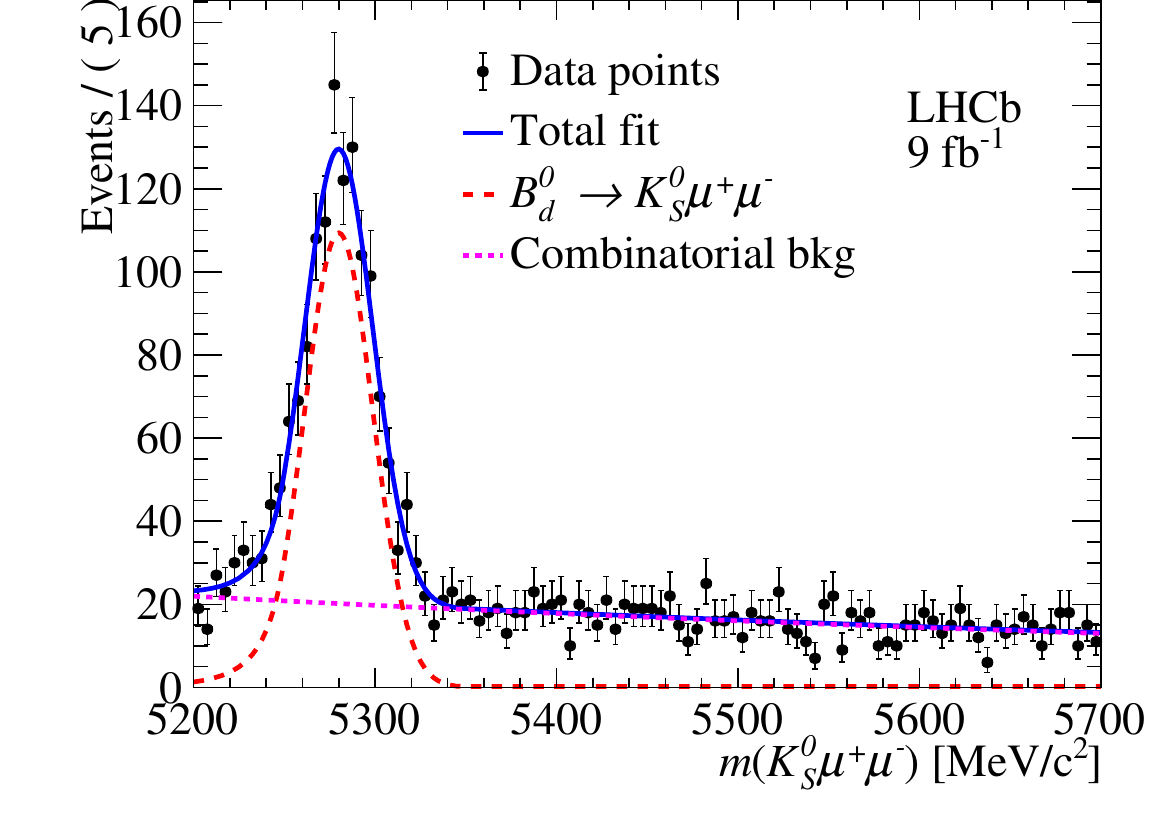}  
   \caption{
   Mass distribution of the selected \decay{\Bd}{\KS \mup\mun} candidates, with the fit result also shown. 
  }
  
  \label{fig1}
  \end{figure}

The mass distribution of the selected \Bd candidates combining all categories is shown in Fig.~\ref{fig1}, together with the combined fit results. 
The total signal yield is  $941 \pm 55$.
All selected candidates are then sorted into two groups corresponding to the Run~1 and Run~2 periods, respectively, which are used in the subsequent decay-time fit to determine the \CP violation parameters.

The decay-time distribution  of  \Bd (\Bdb) mesons produced in  $pp$ collisions is described by
\begin{equation}
\label{eq2}
P_{\Bd(\Bzb)}(t) \propto     e^{-t /\tau_{\Bd} } \left[\cosh{\left(\frac{\Delta\Gamma_d t}{2}\right)} \mp S \sin(\Delta m_d t) \pm C \cos(\Delta m_d t) +D\sinh{\left(\frac{\Delta\Gamma_d t}{2}\right)} \right]\,,  
\end{equation}
where $\tau_{\Bd}$ is the lifetime of the \Bd meson. The  upper (lower) sign  corresponds to  \Bz (\Bzb) decays.  
The parameter $D$  parametrizes \CP-violating effects originating from  $\Delta\Gamma_d$. 
Since $\Delta\Gamma_d$ is  compatible with zero~\cite{PDG2024}, the $\cosh{\frac{\Delta\Gamma_d t}{2}}$ term  is approximated to 1, and the $\sinh{\frac{\Delta\Gamma_d t}{2}}$ term in Eq.~\ref{eq2} is neglected in the baseline measurement.
The reconstructed  decay-time distribution is affected by the experimental effects discussed below.

The  flavor of the $\Bd$  or $\Bdb$ meson at production is identified using both the same-side (SS)~\cite{LHCb-PAPER-2016-039} and opposite-side (OS)~\cite{LHCb-PAPER-2011-027} flavor tagging algorithms. The SS algorithm exploits charged particles from the signal $b$-quark fragmentation, such as pions and protons. The OS algorithm utilizes decay products from the associated partner $b$-hadron, including kaons, leptons, reconstructed charm hadrons, and the charge information from its decay vertex. Each algorithm provides a tagging decision $d$ with an estimated mistag probability, where $d = +1,-1$, or $0$ if the meson is classified as \Bd, \Bdb or untagged, respectively.  
The estimated mistag probability is calibrated with a dedicated  software package~\cite{lhcb_ftcalib:2024}, 
using  \decay{\Bu}{\jpsi \Kp} 
and \decay{\Bz}{\jpsi \Kstar(892)^0} decays  for the OS  and SS tagging method,  respectively. 
The calibration samples are weighted to match the distributions in the signal sample of kinematic properties (\pt, $\eta$, and azimuthal angle of the signal \Bd candidates) and event characteristics (PV and track multiplicities). The package yields the calibrated mistag probability $\omega$ for each candidate. 
The OS and SS tagging decisions and mistag probabilities are then combined to give the final tagging decision and mistag probability for each $B$ meson. 
    The total effective tagging efficiency is$\etag$ $\left<D^2\right>  =(2.7\pm1.1)\%$ and $(4.1\pm0.6)\%$ for the Run~1 and Run~2 samples, respectively. Here \etag is the fraction of of $B$ candidates that have a nonzero tagging decision, and $D^2$ is  the effective dilution squared depending on the event-by-event mistag probability~\cite{LHCb-PAPER-2011-027, LHCb-PAPER-2016-039}.

Assuming no production or tagging asymmetries for 
\Bd  and \Bdb mesons, the decay-time distribution becomes
\begin{align}
\label{eq3}
P(t, d, \omega) \propto  \left[1 + d(1 - 2\omega\right)] P_{\Bd}(t) + \left[1 - d(1 - 2\omega\right)] P_{\Bdb}(t)\,.
\end{align}
The decay-time resolution is accounted for by convolving Eq.~\ref{eq3} with a resolution function 
consisting of the sum of three Gaussian functions. Each Gaussian has the same mean and a width depending linearly on  the per-candidate decay-time uncertainty.
    The parameters of the resolution model are obtained by fitting the samples  of  prompt $\KS\mup\mun$ candidates,  which are constructed from a $K_{S}^{0}$ and two muon candidates passing the same selection criteria as that for the signal channel, except the minimum requirement on the  $B$ meson decay time.
The resulting resolution model, with an effective resolution of about 60\fs, affects the \CP-violating parameters by less than $0.1\%$, which is negligible compared to the statistical uncertainties.

The reconstruction, trigger, and selection requirements result in a decay-time-dependent signal efficiency. This effect is accounted for by multiplying the decay-time distribution by an acceptance function parameterized by a cubic spline function~\cite{Karbach-2014qba} with six knots at 0.5, 1.0, 1.5, 2.0, 3.0, 8.0\ps  and eight coefficients. One coefficient is fixed to unity for normalization, and the remaining coefficients are allowed to vary in the fit to data described below.

A weighted unbinned maximum-likelihood fit~\cite{Plan2012,Xie:2009rka} is  simultaneously performed  on the Run~1 and Run~2 data samples to the decay times of the 
\decay{\Bd}{\KS \mup\mun} candidates to determine the \CP violation parameters $S$ and $C$.
In this fit, only the signal component is explicitly modeled while the background contributions are statistically subtracted via the signal weights calculated  with the \sPlot~\cite{Pivk:2004ty} method  using the \Bd mass as the discriminating variable. This method   assumes that the decay time and \Bd mass are uncorrelated in both the signal and individual background components. Potential correlations are treated as systematic uncertainties.

In the fit, the \Bd lifetime and the mass difference between the two  neutral \B   mass eigenstates  are constrained to their world-average values, \mbox{$\tau_{\Bz}= 1.517 \pm 0.004\ps$} and $\dmd = 0.507 \pm 0.002\invps$~\cite{PDG2024}.
This allows the time acceptance  parameters to be  determined entirely from the fit to the signal data. The decay-time resolution  and flavor-tagging calibration parameters are constrained to the fit results obtained in the corresponding calibration  channels.

\begin{figure}[t]
\centering
    
     \includegraphics[width=0.47\linewidth]{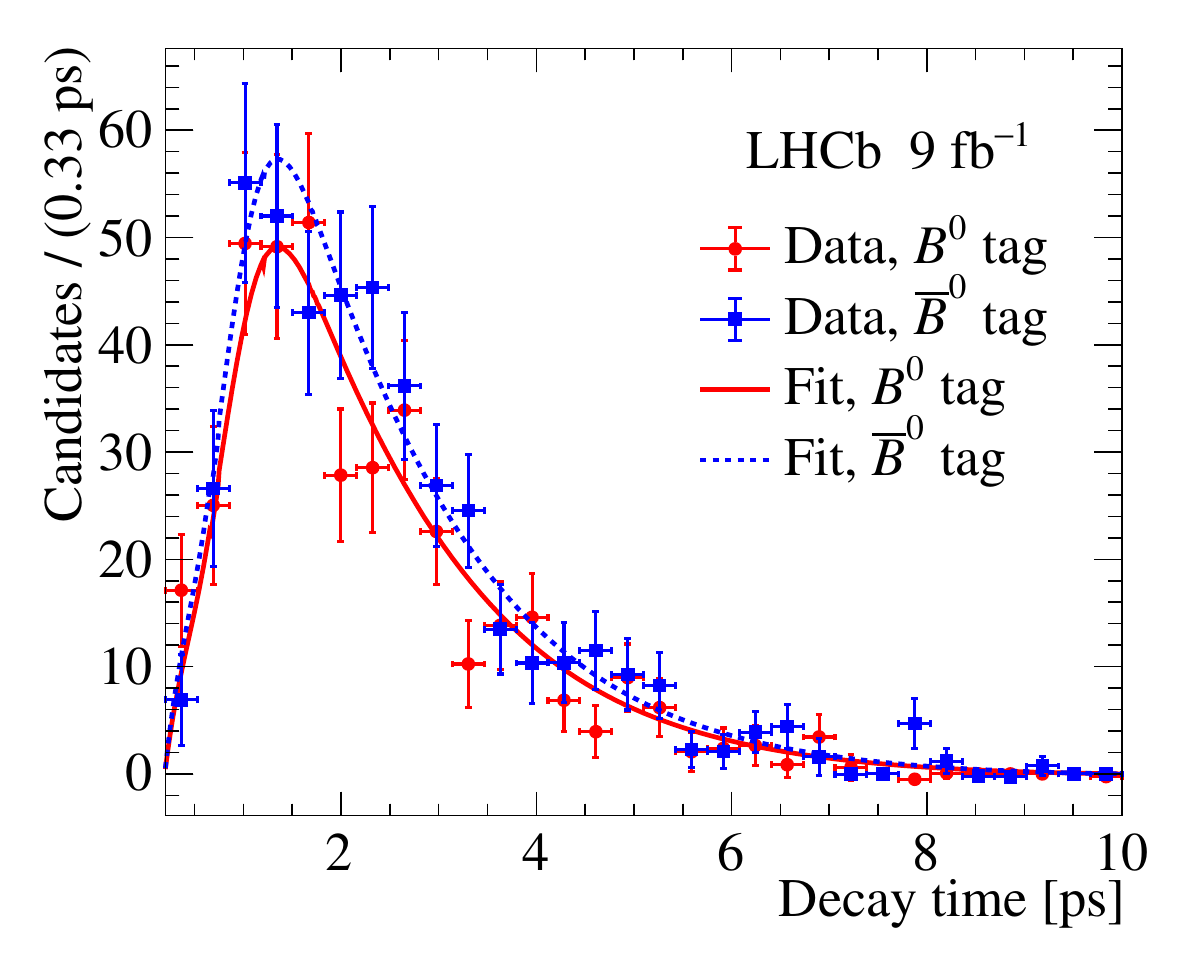}
   \includegraphics[width=0.47\linewidth]{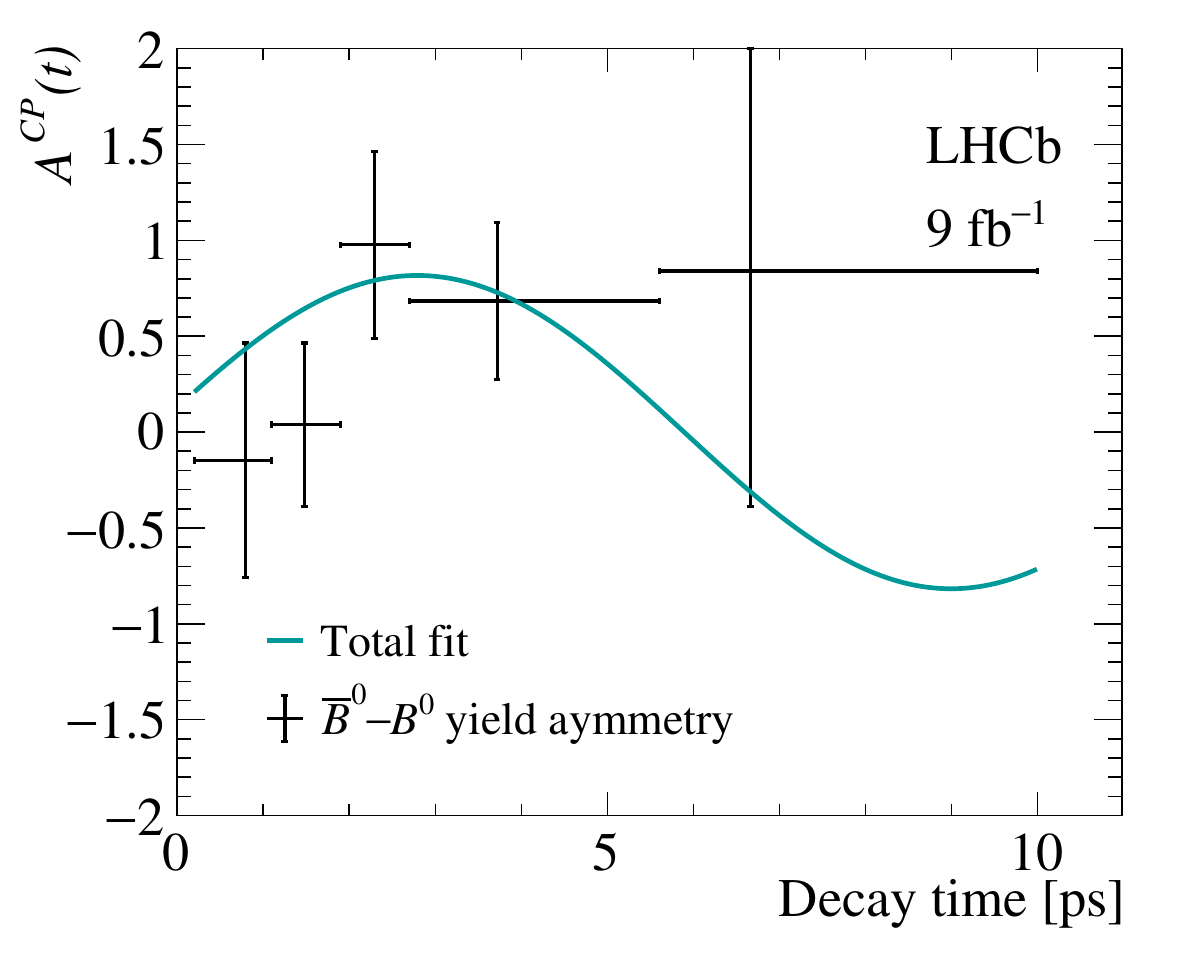}
   \caption{
  Left: background-subtracted decay-time distributions for $\Bdb$ and $\Bd$ candidates identified through flavor tagging, where the  fit result is also shown. The difference between $\Bdb$ and $\Bd$ distributions reveals the \CP asymmetry.  Right:   time-dependent \CP asymmetry. The data points show the binned signal yield asymmetry, with tagging dilution effects corrected for. The result of the maximum-likelihood fit is shown as the solid curve.
  }
  \label{fig:decaytime}
  \end{figure}

\begin{figure}[tb]
\centering
   \includegraphics[width=0.6\linewidth]{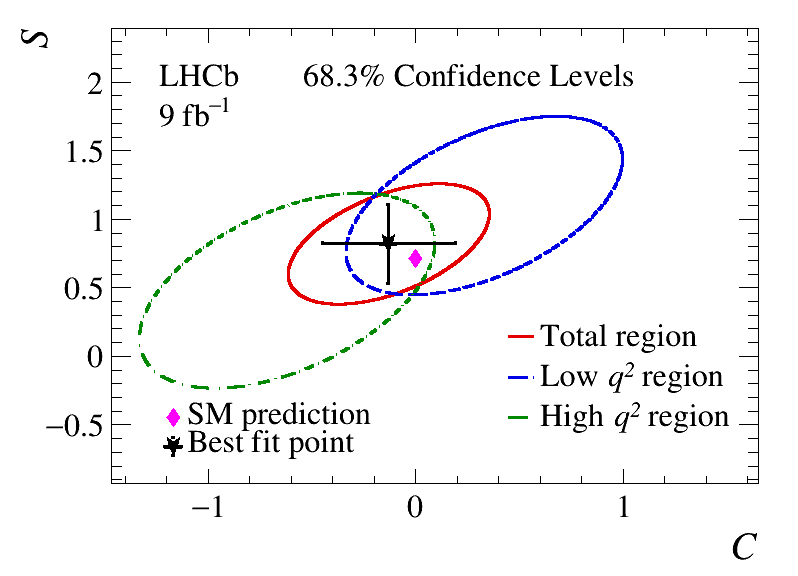}
   \caption{
 Two-dimensional confidence regions for the  parameters $C$ and
$S$ from profile likelihood scans,  only statistical  uncertainties are considered in this plot. The green dash-dotted, blue dashed and red solid contours represent the $68.3\%$ confidence regions  for the low-$q^2$,  high-$q^2$ and total regions, respectively.  The black marker with error bars represents the  fit results  in the total sample, while the magenta marker stands for the SM prediction.
  }
  \label{fig:2dscan1}
  \end{figure}

The background-subtracted decay-time distributions of the tagged \Bdb and \Bd signal candidates in the combined data sample are shown in Fig.~\ref{fig:decaytime}, together with the  baseline fit using Eq.~\ref{eq2}. The corresponding time-dependent \CP asymmetry is also presented, which is computed from the tagged yields after correcting for the effects of flavor-tagging dilution.
The \CP violation parameters are determined from the time-dependent fits both for the entire kinematically  allowed $q^2$ range (excluding the $\jpsi$ and $\psi(2S)$ regions) and, to probe potential $q^2$ dependency, in the $q^2$ ranges below and above the \jpsi region.  The results are   summarized in Table~\ref{tab:cp_results}
 and Fig.~\ref{fig:2dscan1}.
The 68.3\%, 95.5\% and 99.7\% confidence regions under the physical constraint $C^2+S^2\le 1$ are further obtained  for the total $q^2$ region using the Feldman--Cousins
method~\cite{FCscan,BAKER1984437} and  shown in the End Matter.

\begin{table}[tb]
\centering
\caption{Summary of the \CP violation parameters from the time-dependent fit to \decay{\Bd}{\KS \mup\mun} decays in the total, low  and high $q^2$ ranges, excluding the charmonium regions.Only statistical uncertainties are  considered.} 
\label{tab:cp_results}
\begin{tabular}{lcc}
\hline
$q^2$ range $[\gevgevcccc]$ & $C$ & $S$ \\
\hline
Total 
& $-0.13 \pm 0.32 $ & $+0.82 \pm 0.29 $  \\
Low: $q^2<8.0$ 
& $+0.33 \pm 0.44$ & $+1.10 \pm 0.43$\\
High: $11.0<q^2<15.0$ or $q^2>17.5$
 & $-0.62 \pm 0.47$ & $+0.48 \pm 0.47 $ \\
\hline
\end{tabular}
\end{table}

Several  sources  of systematic uncertainties for the
\CP-violation observables are investigated and listed in Table~\ref{tab:syst_sum}.   
A large ensemble of pseudoexperiments generated according to the baseline fit results is used to validate the fit procedure. The observed distributions of the parameters from these pseudoexperiments demonstrate that the statistical uncertainties are correctly treated. A bias in the $C$ and $S$  parameters is  observed, which is due to the limited data sample size. 
The consistency of the tagging calibration
parameters between the calibration and signal channels 
is tested  using simulated samples, and that between \Bd and \Bdb mesons is investigated  using the calibration channels \decay{\Bu}{\jpsi \Kp} 
and \decay{\Bz}{\jpsi \Kstarz}. The observed differences represent the leading source of systematic uncertainties. 
Systematic  uncertainties associated with the modeling of the signal and background contributions are evaluated by repeating the mass fit with alternative functions, namely the sum of a Student's t-distribution and a Gaussian function for the signal and a second-order polynomial for the combinatorial background.  
Then, the weighted maximum-likelihood fit is performed, and the differences in the resulting parameters from those of the baseline fit are assigned as  systematic uncertainties.
The effect of neglecting a small correlation between the reconstructed \Bd mass  and decay time in the signal weight determination when applying the \sPlot method is studied by reevaluating the signal weights in bins of the decay time and performing the decay-time fit again.  
The effects of a nonzero $\Bz$--$\Bzb$ production asymmetry and a nonzero $\Delta\Gamma_d$ are evaluated by varying their values within their  uncertainties~\cite{LHCb-PAPER-2014-042,PDG2024} and performing  the fits again. The resulting largest  variations in \CP violation parameters are assigned as systematic uncertainties.
The systematic uncertainties associated with  the decay-time acceptance and resolution models are evaluated using splines with different knot numbers and positions and found to be negligible.
The total systematic uncertainties, evaluated as the  quadrature sums of the contributions discussed above,  are
significantly smaller than the statistical ones.

\begin{table}[t!]
\begin{center}
        \caption{Summary of absolute systematic uncertainties for the parameters $C$ and $S$.}
        \label{tab:syst_sum}
        \begin{tabular}{l|cc}
  \hline   
         Sources&$C$ & $S$ \\
\hline
         Fit bias&0.017&0.015\\
         Flavor tagging  &0.031&0.037\\
          Mass fit models  & 0.010&0.011\\
          Mass-time correlation  & 0.013&0.018\\
          Nonzero $\Delta\Gamma_{d}$&0.007&0.005\\
         Production asymmetry&0.006&0.004\\
         Time acceptance and resolution & negligible&negligible\\
        
\hline
         Total&0.041&0.046\\
         \hline
        \end{tabular}
\end{center}
\end{table}

The robustness of the analysis  results is checked 
 by splitting the data sample by \KS reconstruction categories, magnet polarities, and data-taking periods, 
 by varying the \Bd  mass and decay-time ranges, and  by obtaining alternative determination of the  $S$ parameters from the time-integrated \CP asymmetry.
The results are all compatible with each other and with the baseline fit results. As a cross-check, the analysis procedure has also been applied to the control channel  \decay{\Bd}{\jpsi\KS }. The corresponding results are consistent with those reported in Ref.~\cite{LHCb-PAPER-2023-013}.

In conclusion, this Letter reports a measurement of the time-dependent \CP asymmetry in  \mbox{$\Bd\rightarrow \KS \mu^{+}\mu^{-}$ }  decays using data collected by the \lhcb experiment during the Run~1 and Run~2 data-taking periods. 
The \CP violation parameters in the  full $q^2$ range, excluding the dimuon resonant regions due to \jpsi and \psitwos states, are determined to be
\begin{align*}
C &= -0.13 \pm 0.32 \, (\text{stat}) \pm 0.04 \, (\text{syst})\,,\\
S &= +0.82 \pm 0.29 \, (\text{stat}) \pm 0.05 \, (\text{syst}) \,,
\end{align*}
 with a statistical correlation coefficient of 0.50.
  This is the first measurement of time-dependent \CP violation in \decay{b}{s \ell^+ \ell^-} transitions, and the results are consistent with  the SM prediction.  
This study  demonstrates a new avenue to probe  physics beyond the SM  in  \decay{b}{s \ell^+ \ell^-} transitions, complementary to measurements of branching fractions, angular observables and lepton-flavor universality observables.

\clearpage
\section*{End Matter}
\setcounter{section}{0}
\section{Confidence regions  within the physical boundary}

\begin{figure}[h]
\centering
    \includegraphics[width=0.5\linewidth]{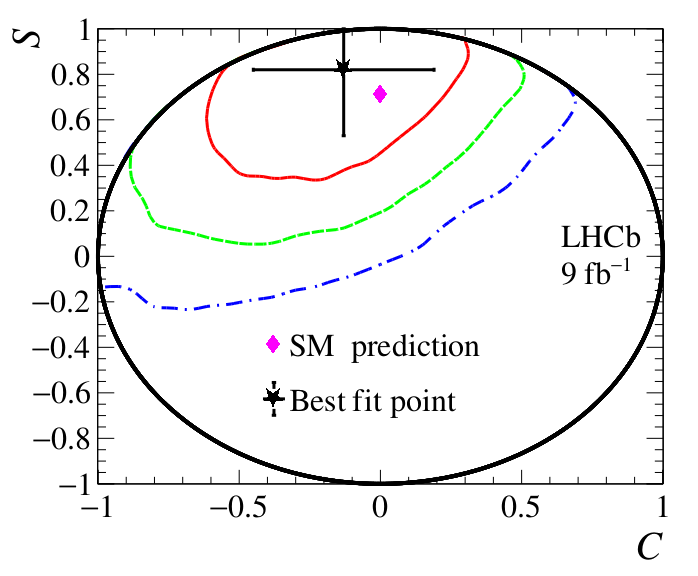}\\
\caption{Two-dimensional confidence regions for $C$ and
$S$ based on a likelihood-ratio ordering technique, corresponding to the total $q^2$ range. The red solid, green dashed and blue dash-dotted contours represent the 68.3\%,
95.5\%, and 99.7\% confidence intervals given the physical constraint $C^2 + S^2 \leq 1$. The black marker with error bands represents the  fit results  for $C$ and $S$.  The magenta diamond marker stands for the SM prediction.  Only statistical uncertainties are considered in this plot.
}
  \label{fig:2dscan2}
  \end{figure}

The results shown in Fig.~\ref{fig:2dscan1} are obtained without applying any physical constraint on the \CP violation parameters. In fact, the normalization of the  coefficients of the time-dependent decay rate requires \(C^2 + S^2 \leq 1\).  Confidence intervals satisfying this requirement are   derived using a constrained likelihood-ratio test. Specifically, the maximum-likelihood estimation and the ordering of parameter points are performed within the restricted domain defined by this inequality. The results are presented in Fig.~\ref{fig:2dscan2}. Systematic uncertainties are excluded from this evaluation as their contribution is negligible compared to the statistical uncertainties.

\clearpage

\clearpage

\section*{Acknowledgements}
%
%
\noindent We express our gratitude to our colleagues in the CERN
accelerator departments for the excellent performance of the LHC. We
thank the technical and administrative staff at the LHCb
institutes.
We acknowledge support from CERN and from the national agencies:
ARC (Australia);
CAPES, CNPq, FAPERJ and FINEP (Brazil); 
MOST and NSFC (China); 
CNRS/IN2P3 and CEA (France);  
BMFTR, DFG and MPG (Germany);
INFN (Italy); 
NWO (Netherlands); 
MNiSW and NCN (Poland); 
MCID/IFA (Romania); 
MICIU and AEI (Spain);
SNSF and SER (Switzerland); 
NASU (Ukraine); 
STFC (United Kingdom); 
DOE NP and NSF (USA).
We acknowledge the computing resources that are provided by ARDC (Australia), 
CBPF (Brazil),
CERN, 
IHEP and LZU (China),
IN2P3 (France), 
KIT and DESY (Germany), 
INFN (Italy), 
SURF (Netherlands),
Polish WLCG (Poland),
IFIN-HH (Romania), 
PIC (Spain), CSCS (Switzerland), 
GridPP (United Kingdom),
and NSF (USA).  
We are indebted to the communities behind the multiple open-source
software packages on which we depend.
Individual groups or members have received support from
RTP (Australia), 
Key Research Program of Frontier Sciences of CAS, CAS PIFI, CAS CCEPP (China); 
Minciencias (Colombia);
EPLANET, Marie Sk\l{}odowska-Curie Actions, ERC and NextGenerationEU (European Union);
A*MIDEX, ANR, IPhU and Labex P2IO, and R\'{e}gion Auvergne-Rh\^{o}ne-Alpes (France);
Alexander-von-Humboldt Foundation (Germany);
ICSC (Italy); 
Severo Ochoa and Mar\'ia de Maeztu Units of Excellence, GVA, XuntaGal, GENCAT, InTalent-Inditex and Prog.~Atracci\'on Talento CM (Spain);
the Leverhulme Trust, the Royal Society and UKRI (United Kingdom).





\addcontentsline{toc}{section}{References}
\bibliographystyle{LHCb/LHCb}
\bibliography{main,LHCb/standard,LHCb/LHCb-PAPER,LHCb/LHCb-CONF,LHCb/LHCb-DP,LHCb/LHCb-TDR}

\newpage
\centerline
{\large\bf LHCb collaboration}
\begin
{flushleft}
\small
R.~Aaij$^{38}$\lhcborcid{0000-0003-0533-1952},
A.S.W.~Abdelmotteleb$^{58}$\lhcborcid{0000-0001-7905-0542},
C.~Abellan~Beteta$^{52}$\lhcborcid{0009-0009-0869-6798},
F.~Abudin\'en$^{60}$\lhcborcid{0000-0002-6737-3528},
T.~Ackernley$^{62}$\lhcborcid{0000-0002-5951-3498},
A. A. ~Adefisoye$^{70}$\lhcborcid{0000-0003-2448-1550},
B.~Adeva$^{48}$\lhcborcid{0000-0001-9756-3712},
M.~Adinolfi$^{56}$\lhcborcid{0000-0002-1326-1264},
P.~Adlarson$^{88}$\lhcborcid{0000-0001-6280-3851},
C.~Agapopoulou$^{14}$\lhcborcid{0000-0002-2368-0147}
% ...

{\footnotesize \it

% ...
$^{14}$Universit{\'e} Paris-Saclay, CNRS/IN2P3, IJCLab, Orsay, France\\
% ...
$^{38}$Nikhef National Institute for Subatomic Physics, Amsterdam, Netherlands\\
% ...
$^{48}$Instituto Galego de F{\'\i}sica de Altas Enerx{\'\i}as (IGFAE), Universidade de Santiago de Compostela, Santiago de Compostela, Spain\\
% ...
$^{52}$Physik-Institut, Universit{\"a}t Z{\"u}rich, Z{\"u}rich, Switzerland\\
% ...
$^{56}$H.H. Wills Physics Laboratory, University of Bristol, Bristol, United Kingdom\\
% ...
$^{58}$Department of Physics, University of Warwick, Coventry, United Kingdom\\
% ...
$^{60}$School of Physics and Astronomy, University of Edinburgh, Edinburgh, United Kingdom\\
$^{61}$School of Physics and Astronomy, University of Glasgow, Glasgow, United Kingdom\\
$^{62}$Oliver Lodge Laboratory, University of Liverpool, Liverpool, United Kingdom\\
% ...
$^{70}$Syracuse University, Syracuse, NY, United States\\
% ...
$^{88}$Department of Physics and Astronomy, Uppsala University, Uppsala, Sweden, associated to $^{61}$\\
% ...
}
\end{flushleft}
\end{document}